\documentclass[letter,seceqn,secthm]{elsart}

\usepackage{latexsym}   
\usepackage{amssymb}
\usepackage{amsmath}
\usepackage{graphicx}
\newcommand{\goth}{\mathfrak}
\newcommand{\eref}[1]{Eq.~(\ref{#1})}
\newcommand{\sref}[1]{Section~\ref{#1}}
\newcommand{\fref}[1]{Fig.~\ref{#1}}

\newcommand{\Eref}[1]{Equation~(\ref{#1})}
\newcommand{\Sref}[1]{Section~\ref{#1}}
\newcommand{\Fref}[1]{Figure~\ref{#1}}

\newcommand{\abs}[1]{\lvert#1\lvert}
\newcommand{\rmd}{\mathrm{d}}
\newcommand{\rme}{\mathrm{e}}

\newenvironment{rmk}[1][Remark. ]{\textit{#1 \/}}{}

\begin{document}

\begin{frontmatter}
\title{Intersecting Loop Models on ${\mathbb Z}^d$: Rigorous Results}

\author{L. Chayes}

\address{Department of Mathematics, UCLA, Los Angeles, CA 90095-1555,
  USA}

\author{Leonid P. Pryadko} 

\address{School of Natural Sciences, IAS,
  Princeton, NJ 08540, USA}

\author{Kirill Shtengel\thanksref{corresp}}

\address{Department of Physics, UCLA, Los Angeles, CA 90095-1547,
  USA}

\thanks[corresp]{Corresponding author (E-mail:
  \texttt{shtengel@physics.ucla.edu})}

\begin{abstract}
  We consider a general class of (intersecting) loop models in $d$
  dimensions, including those related to high-temperature expansions
  of well-known spin models. We find that the loop models exhibit some
  interesting features -- often in the ``unphysical'' region of
  parameter space where all connection with the original spin
  Hamiltonian is apparently lost. For a particular $n=2$, $d=2$ model,
  we establish the existence of a phase transition, possibly
  associated with divergent loops.  However, for $n \gg 1$ and
  arbitrary $d$ there is no phase transition marked by the appearance
  of large loops.  Furthermore, at least for $d=2$ (and $n$ large) we
  find a phase transition characterised by broken translational
  symmetry.
\end{abstract}


\end{frontmatter}

\section{Introduction}
In the recent years there has been much interest in various loop
models.  Loop models are graphical models defined by drawing closed
loops along the bonds of the underlying lattice. The loops may come in
$n$ different flavours (colours).  No two loops can share a bond,
while sharing a vertex is generally allowed. Explicitly, the bond
configurations are such that each vertex houses an even number --
possibly zero -- of bonds of each colour.  Each loop configuration is
assigned a ``weight'' that depends on the number of participating
vertices of each type.  In the cases of interest these weights are
actually positive hence, at least in finite volume, they define a {\em
  probability measure\/} on the set of all loop configurations.  Thus,
for a finite lattice the loop partition function may be written as:
\begin{equation}
\label{part}
  Z = \sum_{\mathcal{G}}R^{b} \nu_1^{m_1} \nu_2^{m_2} 
  \ldots \nu_V^{m_V} \, ,
\end{equation} 
with the sum running over all allowed loop configurations $\mathcal{
  G}$.  Here $b$ is the total number of participating bonds, $m_i$
($i=1,\ldots,V$) is the number of vertices of type $i$ and $\nu_i$ is
the corresponding vertex factor.\footnote{Many authors consider an
  additional factor of the form $F_1^{l_1} F_2^{l_2} \ldots F_n^{l_n}$
  where $F_i$ is a ``loop fugacity'' and $l_i$ is the number of loops
  of the $i$-th colour.  Although the objects $l_i$ are unambiguous
  when self-intersections are forbidden, in the general case they are
  not easily defined. Nevertheless, the essence of such a term -- at
  least in the case of integer $F_i$ -- is captured by the
  introduction of additional colours.}  This definition is slightly
different from the one typically found in literature ({\em cf.}\ 
Refs.~\cite{Kondev-96:loops,Warnaar-93}) since it also includes
the bond fugacity $R$.  Although strictly speaking it is not needed
(since the bond fugacity can always be incorporated into the vertex
factors), we find it convenient to keep $R$ as a separate parameter.
We remark that by relabeling the empty bonds as an additional colour,
these models may be formally regarded as ``fully packed''.

The reason loop models have been extensively studied is because they
appear quite naturally as representations (often approximate) of
various statistical - mechanical models. These include, among others,
the Ising model (this approach dates back to Kramers and
Wannier~\cite{KW} and was later used to solve the model
exactly~\cite{KacWard,Vdov}), the Potts model (polygon
expansion~\cite{Baxter}), $\mathrm O(n)$ spin models~
\cite{Domany-81,Nienhuis:On,Nienhuis:On_square,BN:On,BatNW:On}, 1-D
quantum spin models~\cite{Aizenman-94}, a supersymmetric spin
chain~\cite{SuperSpin}, the $q$-colouring
problem~\cite{Baxter:col,Kondev:4col} and polymer
models~\cite{JK-98,JK-99}.

Here we consider the loop models explicitly related to the
high-temperature expansions of the standard $\mathrm O(n)$,
corner-cubic (AKA diagonal-cubic) and face-cubic spin models. This is,
in fact, the same set of models that was treated in
Ref.~\cite{Domany-81}.  However, in this paper, we provide a careful
treatment of the large $n$ cases -- and we treat the standard
$d$-dimensional lattices.  As a result, we arrive at quite unexpected
results concerning the behaviour of these models in the high fugacity
region.

In particular, despite the considerable attention the subject has
received, most authors (with certain exceptions, e.g.\ 
\cite{KW,KacWard,Vdov,Nienhuis:On_square,SuperSpin}) chose to consider
models where only loops of {\em different\/} colours are allowed to
cross each other (if at all). On the other hand spin systems (in the
high-temperature approximation) naturally generate self-intersecting
loops.  In order to avoid this issue, an exorbitant amount of work has
been done on lattices with coordination number $z=3$ (e.g.\ the
honeycomb lattice), where loop intersections simply cannot occur.
Overall this approach appears to be justified since one is usually
interested in the critical properties of the underlying spin systems.
Indeed, consider the archetypal $n$-component spin system with
$\abs{\mathbf S_i} \equiv 1$ and let us write $\exp \bigl(\lambda
\sum_{\langle i, j \rangle} \mathbf{ S}_i \!  \cdot \mathbf{ S}_j
\bigr) \sim \prod_{\langle i, j \rangle} \left(1 + \lambda \mathbf{
    S}_i \! \cdot \mathbf{ S}_j \right)$.  Although as a spin system
the right hand side makes strict sense only if $\abs{\lambda} \leq 1$
(the ``physical regime''), the associated loop model turns out to be
well defined for all $\lambda$.  Since the systems can be identified
for $\abs{\lambda} \ll 1$ it can be argued that the critical
properties of the spin system and those of the loop model are the same
and are independent of the underlying lattice.

Notwithstanding, for $n \gg 1$ any phase transition in the actual spin
system is not anticipated until temperatures of order $1/n$ (i.e.
$\lambda \sim n$), which we note is well outside the physical regime
of the loop model. At first glance this appears to be borne out: the
natural parameter in the loop model (as well as in the spin system)
seems to be $\lambda/ n$. Thus, the loop model could, in principle,
capture the essential features of the spin system up to -- and
including -- the critical point.

We have found such a picture to be overly optimistic.  Indeed,
depending on the specific details, e.g.\ the lattice structure, there
may be a phase transition in the region $1\ll \abs{\lambda} \ll n$
(specifically, $\lambda \sim n^{3/4}$), well outside the physical
regime but well before the validity of the approximation was supposed
to break down.  Furthermore, it would seem that both the temperature
scale and the nature of the transition (not to mention the existence
of the transition) depend on such details.  Finally, we shall
demonstrate that in contrast to their spin system counterparts, the
large-$n$ models have {\em no\/} phase transition -- for any value of
bond fugacity -- associated with the formation of large loops (i.e.\ 
divergent loop correlations).

The structure of this paper is as follows. \Sref{sec:models} is
dedicated to the description of the spin models and their connection
to the loop models.  Specific results for those models with the
two-dimensional spin variable ($n=2$) are presented in \Sref{sec:n=2}.
Finally, \Sref{sec:n_large} contains the discussion of reflection
positivity as well as some results concerning phase transitions in the
large $n$ case.  \vfill

\section{$n$-component models}
\label{sec:models}

\subsection{$\mathrm O(n)$ model}
\label{sec:On}

Let us start by considering the $\mathrm O(n)$ model on some finite
lattice $\Lambda \subset {\mathbb Z}^d$ {\em defined\/} by the
following {partition function}:
\begin{equation}
  Z = \mathrm{Tr}\prod_{\langle i, j \rangle} 
  \left(1 + \lambda \mathbf{S}_i \cdot \mathbf{S}_j \right)
  \label{O(n)_part}
\end{equation}
with $\mathbf{S}_i \in {\mathbb R}^n$, $\abs{\mathbf{S}_i} = 1$ and
$\mathrm{Tr}$ denoting normalised summation (integration) over all
possible spin configurations.  The corresponding loop model is readily
obtained along the lines of a typical ``high-temperature'' expansion.
We write $\mathbf{S}_i \cdot \mathbf{S}_j = S_i^{(1)} S_j^{(1)} +
\ldots + S_i^{(n)} S_j^{(n)}$ and define $n$ different colours (each
associated with a coordinate direction of the $\mathrm O(n)$-spins).
Expanding the product, we have $n$ choices for each bond plus a
possibility of a vacant bond.  Thus, various terms are represented
by $n$-coloured bond configurations: $\mathcal{G} = (\mathcal{G}_1,
\ldots, \mathcal{G}_n)$ with $\mathcal{G}_\ell$ denoting those bonds
where the term $S_i^{(\ell)} S_j^{(\ell)}$ has been selected. Clearly,
the various $\mathcal{G}_\ell$'s are pairwise (bond) disjoint. Thus,
for each $\mathcal{G}$ we obtain the weight
\begin{equation} 
  W_\mathcal{G} = \mathrm{Tr} \prod_{\langle
    i, j \rangle \in \mathcal{G}_1} \lambda S_i^{(1)} S_j^{(1)} \ldots
  \prod_{\langle i, j \rangle \in \mathcal{G}_n} \lambda S_i^{(n)}
  S_j^{(n)}\,.
  \label{bond_weight}
\end{equation}
On the basis of elementary symmetry considerations it is clear that
$W_\mathcal{G} \neq 0$ if and only if each vertex houses an even
number (which could be zero) of bonds of each colour. Once this
constraint is satisfied, we get an overall factor of
$\lambda^{b(\mathcal{G})}$ -- with $b(\mathcal{G})$ being the total
number of participating bonds -- times the product of the {\em vertex
  factors\/} obtained by performing the appropriate $\mathrm O(n)$
integrals.  The details and results of these calculations are
presented in Appendix~\ref{sec:KMF}.  In general, it is seen that the
vertex factors depend only on the number of participating colours and
the number of bonds of each colour emanating from a given vertex,
i.e.\ not on the particular colours that were involved nor on the
directions of these bonds.

In the case of a square lattice ($d=2$) we have only three main types
of (non-empty) vertices: those where two bonds of the same colour join
together, those with two pairs of bonds of two different colours and
those with four bonds of the same colour. These have weights of $1/n$,
$1/n(n+2)$ and $3/n(n+2)$ correspondingly.  Rescaling the bond
fugacity from $R=\lambda$ to $R=\lambda/{n}$ we arrive at the vertex
weights $\nu_1=1$, $\nu_2 =n/(n+2)$ and $\nu_3 = 3n/(n+2)$.

\begin{figure}[hbt]
  \begin{center}
    \includegraphics[width=12cm]{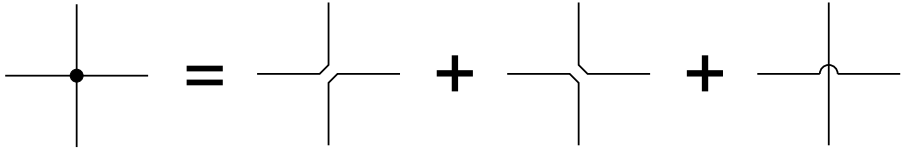}
    \caption{
      Decomposition of a type 3 vertex into three new vertices in the
      two-dimensional O(n) loop model.}
    \label{O(n)_vertices}
  \end{center}
\end{figure}

The factor of 3 relating $\nu_2$ to $\nu_3$ has an interesting
interpretation (which, as shown in Appendix~\ref{sec:KMF}, turns out
to be quite general). Indeed, each vertex of the third type may be
decomposed into three different vertices as shown in
\fref{O(n)_vertices}. Each of the new vertices is now assigned equal
weight, which is also that of $\nu_2$. We thus split each
$\mathcal{G}$ into $3^{m_3}$ different graphs -- each of equal weight
-- in which every vertex with four bonds now provides explicit
instructions relating outgoing and incoming directions of an
individual walk.

Hence in every such graph (now defined with the walking instructions
encoded at every vertex) the individual loops are well defined.
Furthermore, changing the colour of any loop does not change the
weight of the graph.  Thus we may write
\begin{equation}
  Z = \sum_{\mathcal{K}} 
  \left(\frac{\lambda}{n}\right)^{b(\mathcal{K})}
  \left(\frac{n}{n+2}\right)^{m(\mathcal{K})} 
  n^{\ell(\mathcal{K})}
  \label{O(n)_loops}
\end{equation}
where the summation now takes place over all configurations
$\mathcal{K}$ of colourless loop graphs in which every vertex
housing four bonds is resolved by ``walking instructions'', and $\ell$
is the number of such loops (being now defined completely
unambiguously).  In addition to the advantages of a manifestly
colourless expression, the above permits continuation to non-integer
$n$.

\begin{figure}[htb]
  \begin{center}
    \includegraphics[width=13cm]{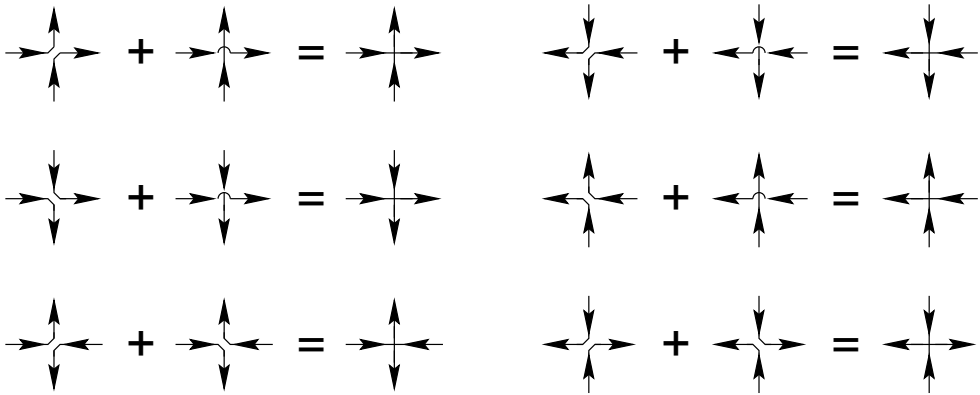}
    \caption{
      Mapping of a fully packed $\mathrm O(2)$ model onto the square ice
      model.}
    \label{6-vertex}
  \end{center}
\end{figure}

We conclude this subsection with the following series of remarks and
observations.
\begin{itemize}
\item{As shown in Appendix~\ref{sec:KMF}, such vertex decomposition
    works, in fact, for an arbitrary lattice in an arbitrary number of
    spatial dimensions (with the proper weights for vertices housing
    6, 8, etc. bonds).}
\item{Notice that only $\abs{\lambda} \leq 1$ region of the parameter
    space is ``physical'' (in a sense of the underlying Hamiltonian:
    $-\beta H=\sum_{\langle i, j \rangle}\ln [1 + \lambda \mathbf{
      S}_i \cdot \mathbf{ S}_j ]$), while for $\abs{\lambda} > 1$ one
    presumes, no spin Hamiltonian can be written at all. The
    corresponding loop model, however, makes perfect sense in the
    entire parameter space.}
\item{If we consider a 2D XY model ($n=2$), we notice that the factor
    $2^{l(\mathcal{ G'})}$ in \eref{O(n)_loops} can be obtained by
    assigning directions to the colourless loops. The above
    decomposition of type 3 vertices makes this procedure unambiguous.
    Having done that, we can turn this model into a random surface
    model by assigning heights to the plaquettes in such a way that a
    plaquette to the right of a directed bond is always one step
    higher than the plaquette to the left.  Not surprisingly, this
    random surface model turns out to be identical to the one obtained
    by the standard means of Fourier-transforming the original weights
    in \eref{O(n)_part} ({\em cf.}\ 
    Refs.~\cite{JKKN-77,Knops-77:XY-SOS}).}
\item{Finally, it is worth mentioning that in the fully packed limit
    $R \to \infty$, the $n=2$ loop model on the square lattice (with
    the colour degrees of freedom being replaced by assigning
    directions to the loops) turns out to be nothing but the square
    ice model (i.e. the six-vertex model with all six weights being
    equal -- see Ref.~\cite{Baxter} for a definition of this model).
    The mapping between the vertices of these models is shown in
    \fref{6-vertex}.  We remark that the perspective of the ice model
    (and, for that matter, other six-vertex models) as a two colour
    loop model provides additional flexibility in the analysis of
    these systems.  These issues will be pursued in a future
    publication.}

\end{itemize}

\subsection{Corner-cubic model}
\label{sec:corner-cubic}

We now consider the following ``discretised'' modification of the
above $\mathrm O(n)$ model (given by \eref{O(n)_part}):
\begin{equation}
  Z = \mathrm{Tr}\prod_{\langle i, j \rangle} 
  \left[1 + \frac{\lambda}{n} \left({\sigma_i^{(1)}\sigma_j^{(1)}+
  \sigma_i^{(2)}\sigma_j^{(2)}\ldots+
  \sigma_i^{(n)}\sigma_j^{(n)}}\right)\right]
\label{corner_part}
\end{equation}
with $\sigma_i^{(k)} = \pm 1$.  For small values of $\lambda$ this
model may be viewed as a high-temperature limit of a corner-cubic
model. Indeed, it describes an interaction of the type in
\eref{O(n)_part} where spins $\mathbf{ S}_i$ are allowed to point at
the corners of an $n$-dimensional hypercube (with the origin being
placed at the centre of the cube).

Mapping it onto an $n$-colour loop model is almost identical to the
$\mathrm O(n)$ case, with the only difference being the vertex factor:
$\left({\sigma_i^{(1)}}\right)^{2k_1} \ldots
\left({\sigma_i^{(n)}}\right)^{2k_n} = 1 $.  We can choose to
associate the weight of $R = {\lambda}/{n}$ with each bond, thus
making all vertex weights $\nu_i$ to be equal to unity. In other
words, the resulting loops in this model do not interact with each
other via vertices (there is still a hard core bond repulsion,
however). The partition function is then simply
\begin{equation}
  Z = \sum_{\mathcal{G}} 
  \left(\frac{\lambda}{n}\right)^{b(\mathcal{ G})}.
\label{bonds-corner}
\end{equation}
\begin{figure}[htb]
  \begin{center}
    \includegraphics[width=13cm]{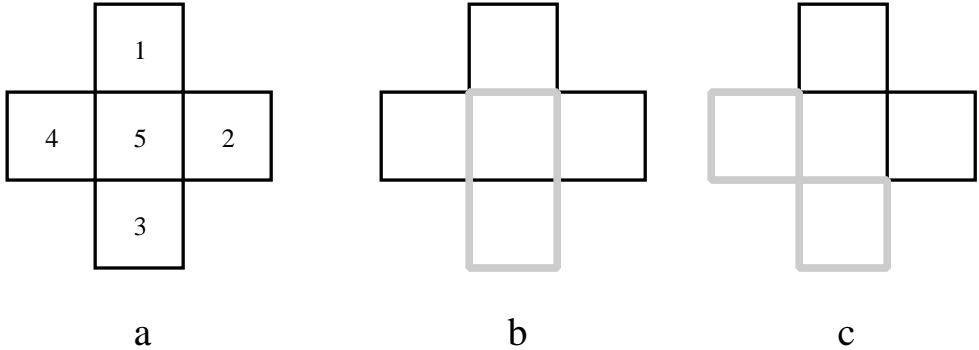}
    \caption{
      A fragment of a possible two-dimensional loop configuration for
      an intersecting loop model. Three different possible colourings
      (out of the total of 32 in the case of $n=2$) for a loop model
      of {\em corner-cubic\/} type are represented here as (a), (b)
      and (c). For the {\em face-cubic\/} type, only monotonous
      clusters like (a) remain allowed.}
    \label{cross-cluster}
  \end{center}
\end{figure}

\subsection{Face-cubic model}
\label{sec:face-cubic}

Finally, let us examine a different model with cubic symmetry given by
the following partition function:
\begin{equation}
  Z = \mathrm{Tr}\prod_{\langle i, j \rangle}
  \left[1 + {\lambda}\left( u_i^{(1)} u_j^{(1)} + u_i^{(2)} u_j^{(2)}
  \ldots + u_i^{(n)} u_j^{(n)}\right)\right]\,. 
  \label{face-cubic}
\end{equation}
Here $u_i^{(k)} = 0,\pm 1$, and for a given site $i$ exactly one of
$u_i^{(k)}$ ($k = 1, 2, \ldots n$) has a non-zero value. In fact, one
may think of $u$'s as components of an $n$-dimensional unit vector
that is only allowed to point along the coordinate axes (or from the
centre to the faces of an $n$-dimensional hypercube - thus the name
face-cubic). While the corner-cubic model described earlier had $2^n$
degrees of freedom per site (the number of corners of a hypercube),
the present model has only $2n$ such degrees of freedom (the number of
faces).

Once again, the corresponding loop model is obtained by performing
multiplication in \eref{face-cubic} and then summing the resulting
terms over all possible values of $u$'s. But since for each site $i$
only one of the spin components $u_i^{(k)} \neq 0$ at a time, we
notice that no terms that mix different $k$'s are allowed. In terms of
resulting loops this means hard-core repulsion of different colours:
only loops of the same colour can share a vertex.  The vertex factors
are now: zero for any vertex with multiple colours and $1/n$ for vertices
with two or more bonds of the same colour; the bond fugacity is given
by $R = \lambda$.

\section{Results for the $n=2$ case}
\label{sec:n=2}
\subsection{The  $n=2$ models with cubic symmetry, Ashkin--Teller 
  and random surface  models}
\label{sec:AT}

In this section we shall restrict our attention to the models with
cubic symmetries. Firstly, let us slightly change the notations for
convenience: let $\sigma_i^{(1)}=\sigma_i$ and $\sigma_i^{(2)}=\tau_i$
for the corner-cubic, while $u_i^{(1)}=u_i$ and $u_i^{(2)}=v_i$ for
the face-cubic model. The corresponding partition functions are then
written as
\begin{equation}
  Z_\mathrm{CC} = \mathrm{Tr}\prod_{\langle i, j \rangle} 
  \left[1 + \frac{\lambda}{2} \left({\sigma_i \sigma_j +
  \tau_i \tau_j}\right)\right]
\label{CC2}
\end{equation}
and
\begin{equation}
  Z_\mathrm{FC} = \mathrm{Tr}\prod_{\langle i, j \rangle} 
\left[1 + {\lambda}\left( u_i u_j + v_i v_j\right)\right]\,.
\label{FC2}
\end{equation}

While we have seen that the loop models generated by these partition
functions are very different, the spin models themselves turn out to
be identical.  Indeed, \eref{FC2} is obtained from \eref{CC2} by the
following transformation: $u_i = (\sigma_i + \tau_i)/2$, $v_i =
(\sigma_i - \tau_i)/2$. This is equivalent to a $45^\mathrm{o}$
rotation in the spin space (along with a $\sqrt{2}/2$ rescaling) and
is very specific to the $n=2$ case.  In turn, both models are
equivalent to the Ashkin--Teller model~\cite{AT} with a particular
choice of parameters that will be detailed in \Sref{sec:AT_duality}.

The two-colour loop models generated by \eref{CC2} and \eref{FC2} are
given by the following sets of parameters in \eref{part}:
$R=\lambda/2$, with all vertices having weight one, and $R=\lambda$,
with all multi-colour vertices given weight zero and all other
non-empty vertices given weight one-half respectively.

Turning our attention to the particular case of two spatial dimensions,
we remark that in the former model one can sum over all possible
colourings to obtain the following result for the partition function:
\begin{equation}
  Z = \sum_{\mathcal{G'}} \left(\frac{\lambda}{2}\right)^{b(\mathcal{ G'})} 
  2^{f(\mathcal{ G'})}
\label{FK-corner}
\end{equation}
with ${b(\mathcal{ G'})}$ being the total number of occupied
(colourless) bonds and ${f(\mathcal{ G'})}$ being the total number of
{\em faces\/} in the clusters they form.  The number of faces is the
minimum number of bonds that one must remove in order for the
remaining clusters to be tree-like.  For example, the cluster in
\fref{cross-cluster} has five {\em faces\/}, while it can at most
consist of four {\em loops\/}. Curiously enough, this result appears
to have no simple generalisation for $n>2$.

It appears that the two-dimensional loop model derived from the
corner-cubic model can not be mapped directly onto a random surface
model.  However, the other loop representation (the one obtained via
expansion of the face-cubic model) does correspond to a random surface
model.  Indeed, consider the following ``recipe'': take a loop
configuration generated by a particular term in the expansion of
\eref{FC2}. Let red be the colour of loops originating from $u$'s,
while blue corresponds to $v$'s.  Take all plaquettes at the outermost
region to be at height zero (these plaquettes are said to form a
substrate).  On this substrate we have clusters of loops. The
outermost boundaries of these clusters are themselves closed loops.
The plaquettes immediately adjacent to these boundaries are assigned
the hight of $+1$ if the loop forming a boundary is red, or $-1$ if it
is blue. These plaquettes, along with any other plaquette accessible
from them without crossing coloured bonds are said to form a plateau
and thus have the same height.\footnote{The strict definition is as
  follows: plaquettes A and B are said to belong to the same plateau
  if and only if there exits an unbroken path along the bonds of a
  {\em dual\/} lattice that connects the centre of plaquette A to the
  centre of plaquette B without crossing a single coloured bond of the
  direct lattice.}  Inside such a plateau region there may be other
loop clusters, that may or may not touch the boundary of a plateau
(only corners are allowed to touch, since no bond sharing between the
loops is possible). Every such cluster is now treated in the same way:
its boundary defines the ``secondary'' plateau with the height being
that of the ``primary'' plateau $\pm 1$ depending on the colour of the
boundary. This procedure is repeated until all plaquettes are assigned
their heights.  As an example, consider the cluster in
\fref{cross-cluster}(a), which may now only consist of the bonds of a
single colour. If this were a red cluster, the heights would be $+1$
for the plaquettes 1-4 and $+2$ for the plaquette 5.

In fact, this description is essentially identical to that given in
Ref.~\cite{Abraham-88} in the context of wetting transition with the
only difference that we allow for two-coloured clusters instead of
single-coloured, and therefore the heights in our case may be both
positive and negative.\footnote{ This random surface model, however,
  has a few important differences with those of a more conventional
  type (like the one obtained for the $\mathrm O(2)$ case). Firstly,
  due to an irreducible four-leg vertex factor, it cannot be described
  by a nearest-neighbour Hamiltonian (i.e. a Hamiltonian that depends
  only on the height difference of neighbouring plaquettes). Secondly,
  since no directions are assigned to the loops separating the
  plateaux, there is no way of deciding on the {\em sign\/} of their
  relative height difference without going through the necessary
  construction steps starting from the {\em outside\/}. By contrast,
  the $\mathrm O(2)$-related random surface model can be constructed
  starting from {\em any\/} plaquette -- a particular choice simply
  determines the overall additive constant.  In this sense the mapping
  between the present random surface and the loop model is {\em
    nonlocal\/}.}

The important feature of this random surface model is that it must
have a phase transition whenever the underlying Ashkin--Teller model
undergoes a transition.

\subsection{Random cluster representation}

Let us derive yet another graphical representation for the $n=2$
(Ashkin--Teller) model considered in the previous section, this time it
will be a {\em random cluster\/} representation closely resembling the
FK representation for the Potts model.  We start from
\eref{corner_part} (with $n=2$), and with the help of the identity
${\sigma_i} {\sigma_j} = 2 \delta_{{\sigma_i} {\sigma_j}} - 1$ rewrite
it as follows:
\begin{equation}
  Z \propto \sum_{\sigma}\prod_{\langle i, j \rangle} 
  \left[1 + {v} \left({\delta_{{\sigma_i} {\sigma_j}} +
\delta_{{\tau_i} {\tau_j}}}\right)\right]
\label{AT_part}
\end{equation}
with $v = {\lambda}/{(1-\lambda)}$.  The random cluster representation
is generated by evaluating the product over all bonds in
\eref{AT_part} and then summing over the possible values of
$\sigma$'s and $\tau$'s. If we think of the bonds originating from the
$\sigma$ variables as green (g), and the bonds originating from the
$\tau$ variables as orange (o), then each of the resulting terms in
the partition function can be graphically represented as a collection
of green and orange clusters as well as empty sites. The clusters of
different colours may share sites, but not bonds. Denoting the
configurations of green and orange bonds as $\omega_\mathrm{g}$ and
$\omega_\mathrm{o}$ respectively, we can then write the partition
function as
\begin{equation}
  Z \propto 
\sum_{\omega} {v}^{b({\omega_\mathrm{g}})}
{v}^{b({\omega_\mathrm{o}})}
2^{c({\omega_\mathrm{g}})}2^{c({\omega_\mathrm{o}})}
\label{AT_FK}
\end{equation}
with ${b({\omega})}$ being the total number of bonds (of a specified
colour), and ${c({\omega})}$ being the number of corresponding
connected components.  The rule for counting the connected components
is as follows: every site that is not a part of a green cluster is
considered to be a separate connected component for the purposes of
${c({\omega}_\mathrm{g})}$, even if this site is a part of an orange
cluster, and vise versa. In particular, the quantities
${v}^{b({\omega_\mathrm{g}})}2^{c({\omega_\mathrm{g}})}$ and
${v}^{b({\omega_\mathrm{o}})}2^{c({\omega_\mathrm{o}})}$ are to be
interpreted exactly as in the usual random cluster models.

\subsection{Self-duality and criticality at $\lambda = 1$}

The duality relations for such random cluster representation of the
standard AT model in two dimensions were established in
Refs.~\cite{CM1} and \cite{PfV}.  Firstly, let us write the generic AT
Hamiltonian as
\begin{equation}
  -\beta H = \sum_{\langle i, j \rangle} 
  \left[K \left({\delta_{{\sigma_i} {\sigma_j}} +
        \delta_{{\tau_i} {\tau_j}}}\right) + L{\delta_{{\sigma_i} 
        {\sigma_j}}\delta_{{\tau_i} {\tau_j}}}\right].
  \label{AT_hamlt}
\end{equation} 

The graphical representation for the partition function is then
obtained along the lines of the previous section. The only difference
is that this time double-coloured (i.e. green and orange at the same
time) bonds are also allowed. The graphical weight of a given bond
configuration $\omega$ is
\begin{equation}
  W(\omega) = 
  {A}^{b({\omega_\mathrm{g}}\vee{\omega_\mathrm{o}})}
  {B}^{b({\omega_\mathrm{g}}\wedge{\omega_\mathrm{o}})}
  2^{c({\omega_\mathrm{g}})}2^{c({\omega_\mathrm{o}})}
  \label{AT_random}
\end{equation}
where
\begin{equation}
  A = \rme^K - 1\:\:\:\:\:
  \mbox{ and }\:\:\:\:\:
  B = \frac{\rme^{L+2K} - 2\rme^K + 1}{\rme^K - 1}\,.
  \label{notations}
\end{equation}
Observe that \eref{AT_FK} describes a particular case of this model
provided that $A = v \equiv \lambda / (1 - \lambda)$ while $B=0$.
\label{sec:AT_duality}

The dual model is obtained by placing orange bonds between the sites
of a dual lattice every time when it does not cross a green bond of
the original lattice. Correspondingly, the green bonds on the dual
lattice are dual to the original orange bonds. The duality relations
are given by:
\begin{equation}
  A^* = 2 B^{-1}\:\:\:\:\:
  \mbox{ and }\:\:\:\:\:
  B^* = 2 A^{-1},
  \label{duality}
\end{equation}
And the model becomes self-dual when $AB = 2$. It is therefore
suggestive that our model becomes self-dual at $\lambda = 1$ (or $v =
\infty$). In order to show that it is indeed {\em exactly\/}
self-dual, we shall perform the above duality transformation to the
orange bonds only (${\omega_\mathrm{o}} \to {\omega_\mathrm{o}^*}$),
leaving the green bonds intact. This results in having green bonds on
both original and the dual lattices. The green bonds on the dual
lattice can be then split into those traversal to the original green
bonds and those traversal to the previously vacant bonds, or
symbolically ${\omega_\mathrm{o}^*} = {\Omega_\mathrm{g}} \vee
\Omega_{\varnothing}$. Here ${\omega_\mathrm{o}^*}$ is the
configuration of (green) bonds {\em dual\/} to the orange bonds while
${\Omega_\mathrm{g}}$ and $\Omega_{\varnothing}$ are the
configurations of bonds {\em transversal\/} to the original green and
vacant bonds correspondingly. The corresponding weight is now given
by:
\begin{equation}
  W(\omega) 
  = {v}^{b({\omega_\mathrm{g}})}
  \left({\frac{2}{v}}\right)^{b({\Omega_\mathrm{g}}) 
    + b({\Omega_{\varnothing}})}
  2^{c({\omega_\mathrm{g}})}\,2^{c({\Omega_\mathrm{g}} 
    \vee \Omega_{\varnothing})}.
  \label{mixed_weight}
\end{equation}
We now observe that ${b({\Omega_\mathrm{g}})} = {b({\omega_\mathrm{g}})}$,
and also that $b({\Omega_{\varnothing}}) \to 0$ as $v \to \infty$ (the
original random cluster model becomes fully packed according to
\eref{AT_FK}). Then the weight in this limit becomes simply
\begin{equation}
  W = {2}^{b({\omega_\mathrm{g}})} \, 2^{c({\omega_\mathrm{g}})}
  \, 2^{c({\Omega_\mathrm{g}})}.
  \label{limiting_weight}
\end{equation}

The model described by such weights is manifestly self-dual since
${\omega_\mathrm{g}^*} = {\Omega_\mathrm{o}} \vee \Omega_{\varnothing} \to
{\Omega_\mathrm{o}}$ and ${\Omega_\mathrm{g}^*} = {\omega_\mathrm{o}} \vee
\omega_{\varnothing} \to {\omega_\mathrm{o}}$ as $v \to \infty$, and
there is a symmetry between the green and the orange bonds.

It is tempting to speculate that a phase transition occurs exactly at
the self-dual point. Although this is plausible, it is not the only
possibility. In particular, there may be a phase transition at some
$\lambda_\mathrm{t} < 1$. However, we can say the following: If at
$\lambda=1$ there is no magnetisation (i.e. percolation of green or
orange bonds) then the theorem proved by two of us~\cite{CS:self-dual}
applies; $\lambda=1$ is a critical point in the sense of infinite
correlation length and infinite susceptibility. The only other
possibility is positive magnetisation at $\lambda=1$ which implies a
magnetic transition -- which could be continuous or first order -- at
some $\lambda_\mathrm{t} \leq 1$. (In particular, this is shown to
happen, with a first order transition for the large-$q$ versions of
these models~\cite{BC}). Although we find these alternative scenarios
unlikely, we have, in any case, established the existence of a
transition in this model for some value of $\lambda$ between zero and
one.

\subsection{Speculative remarks on relation to the critical 
  4-state Potts model}

As mentioned above, in our opinion the most likely scenario is that a
phase transition occurs precisely at $\lambda=1$.  The interesting
question then is that of the universality class. Without any
supporting mathematical statements, we suggest that {\em at\/} this
point our model behaves similarly to the 4-state Potts ferromagnet at
its critical point.  In order to substantiate this claim, let us first
recall the random cluster representation for a $q$-state Potts model:
\begin{equation}
  Z = \sum_{\omega} {K}^{b({\omega})}
  q^{c({\omega})}.
  \label{Potts_FK}
\end{equation}
For $q \leq 4$ this model is universally accepted to have a continuous
transition at the self-dual point $K=\sqrt{q}$.  Thus for the $q=4$
model at the self-dual point we have
\begin{equation}
  Z = \sum_{\omega} {2}^{b({\omega})}
  \,4^{c({\omega})}.
  \label{4-state}
\end{equation} 
On the other hand, we can use \eref{limiting_weight} to rewrite
the partition function of our model at $\lambda=1$ as follows:
\begin{equation}
  Z \propto \sum_{\omega} {2}^{b({\omega})} \, 4^{c({\omega})}
  \, 2^{c({\Omega})-c({\omega})}.
  \label{stupid}
\end{equation}
The difference between the two models is in the last factor of
$2^{c({\Omega})-c({\omega})}$ in \eref{stupid}. It is, however,
reasonable to speculate that it can be neglected. Indeed, on average
${c({\Omega})=c({\omega})}$, and therefore one would expect the
typical value of the difference ${c({\Omega})-c({\omega})}$ to be
sublinear in the system size.  By contrast, the individual terms
$c({\Omega})$ and $c({\omega})$ indeed scale linearly with the size
of the system so this correction may be ``unimportant''.

This, however, does not mean that the two models approach the
self-dual point in a similar fashion. In other words, we expect the
exponents associated with the critical point itself (such as $\eta$
and $\delta$) of the two models to be the same, while this needs not
be true for the exponents associated with the {\em approach\/} to the
critical point (such as $\alpha$, $\beta$ and $\nu$). In fact, the
$\lambda=1$ point of our model may well be an edge of a critical
Kosterlitz-Thouless phase in which case the approach exponents would
take on extreme values (zero or infinity).

\section{Reflection positivity and phase transitions
  in the large $n$ limit}
\label{sec:n_large}

\subsection{Reflection positivity}
\label{sec:RP}

This section concerns the reflection positivity property of the loop
models defined by \eref{part} which in turn permits the analysis of
their large $n$ limit.  Let $\Lambda$ denote a $d$--dimensional torus.
Here, and for the remainder of this paper, it will be assumed that the
linear dimensions of $\Lambda$ are all the same, and are of the form
$L = 2^{\ell}$.  We denote by $N = L^{d}$ the number of sites in the
torus.  Let $\mathcal G$ denote the set of all possible loop
configurations on $\Lambda$.  Finally let $\mathcal P$ denote a
hyperplane perpendicular to one of the coordinate axes which cuts
through the bonds parallel to this axis dividing the torus into two
equal parts.  Let $\mathcal{G}_{\mathcal{L}}$ and
$\mathcal{G}_{\mathcal{R}}$ be the bond configurations on the two
sides of the ``cut'', with the bonds intersected by $\mathcal P$
belonging to {\em both\/} sets. Thus $\mathcal{G} =
\mathcal{G}_{\mathcal{L}} \cup \mathcal{ G}_{\mathcal{R}}$, while
$\mathcal G_{\mathcal{P}} \equiv \mathcal{G}_{\mathcal{L}} \cap
\mathcal{G}_{\mathcal{R}}$ contains only the intersected bonds.  We
now define a map $\vartheta_{\!\mathcal{P}}: \mathcal{G}_{\mathcal{L}}
\to \mathcal{G}_{\mathcal{R}}$ such that it simply reflects the
configuration on the ``left'' to that on the ``right''.  Let $f:
\mathcal{G}_{\mathcal{R}} \to {\mathbb R}$ be a function that depends
only on the bond configuration on the right and define
$\vartheta_{\!\mathcal{P}} f({\goth g_{\mathcal{ L}}}) \equiv
f(\vartheta_{\!\mathcal{P}} (\goth g_{\mathcal{L}}))$ for any $\goth
g_{\mathcal{L}} \in \mathcal{G}_{\mathcal{L}}$.  Similarly, we can use
$\vartheta_{\!\mathcal{P}}$ to map $\mathcal{G}_{\mathcal{R}} \to
\mathcal{G}_{\mathcal{L}}$; with this convention
$\vartheta_{\!\mathcal{P}}^2$ is the identity.  A probability measure
$\mu$ on the set $\mathcal{G}$ is said to be {\em reflection
  positive\/} if for every such $\mathcal{P}$ and any functions $f$
and $h$ as described above $\langle f \,\vartheta_{\!\mathcal{P}} f
\rangle_{\mu} \geq 0$ and $\langle h \;\vartheta_{\!\mathcal{P}} f
\rangle_{\mu} = \langle f \;\vartheta_{\!\mathcal{P}} h
\rangle_{\mu}$.

\begin{thm}
  \label{Theorem_4_1}
  The measures $\mu$ determined by the weights in \eref{part} are
  reflection positive on any even $d$-dimensional torus.
\end{thm}
\begin{pf*}{Proof.}
  Let $\Lambda$ denote one such torus and $\mathcal{P}$ denote one of
  the above described planes.  Let $\goth g_{\mathcal{P}} \in
  \mathcal{G}_{\mathcal{P}}$ be a configuration of bonds going through
  this plane.  Assuming that $\mu(\goth g_{\mathcal{P}}) \neq 0$, let
  us consider the measure $\mu(\cdot \mid\goth g_{\mathcal{P}})$.  Our
  claim is that this splits into two measures, which we will call
  $\mu_{\mathcal{L}}(\cdot \mid\goth g_{\mathcal{P}})$ and
  $\mu_{\mathcal{R}}(\cdot \mid\goth g_{\mathcal{P}})$, defined on
  $\mathcal{G}_{\mathcal{L}}$ and $\mathcal{G}_{\mathcal{R}}$ which
  are independent and identical under the reflection
  $\vartheta_{\!\mathcal{P}}$.
  
  Indeed, it is not hard to see that $\mu(\goth g_{\mathcal{ P}}) \neq
  0$ if and only if $\goth g_{\mathcal{P}}$ has an even number of
  bonds of each colour.  In each half of the torus, the endpoints of
  these bonds serve as ``source/sinks'' for bond configurations.  In
  other words, a configuration in, say, $\mathcal G_{\mathcal{L}}$
  must contain lines of the appropriate colour that pair up these
  sources.  But, aside from having to satisfy these ``boundary
  conditions'', the weights are the same as given in \eref{part}.
  These two measures defined accordingly on $\mathcal G_{\mathcal{L}}$
  and $\mathcal G_{\mathcal{R}}$ are the above mentioned
  $\mu_{\mathcal{L}}(\cdot \mid\goth g_{\mathcal{P}})$ and
  $\mu_{\mathcal{R}}(\cdot \mid\goth g_{\mathcal{P}})$ respectively.
  
  It is clear that if $\goth g_{\mathcal{L}}\in
  \mathcal{G}_{\mathcal{L}}$ then
  \begin{equation}
    \mu_{\mathcal{L}}\left(\goth g_{\mathcal{L}}
    \mid\goth g_{\mathcal{P}}\right) =
    \mu_{\mathcal{R}}\left(\vartheta_{\!\mathcal{P}}
    (\goth g_{\mathcal{L}})\mid\goth g_{\mathcal{P}}\right)\,.
    \label{RP1} 
  \end{equation}
  
  Furthermore if $\goth g_{\mathcal{P}}$ is a configuration and $\goth
  g_{\mathcal{L}}$ is any configuration that agrees with $\goth
  g_{\mathcal{P}}$ and has non-zero weight then for every $\goth
  g_{\mathcal{R}} \in \mathcal{G}_{\mathcal{R}}$ we see that
  \begin{equation}
    \mu_{\mathcal{R}}(\goth g_{\mathcal{R}}\mid\goth g_{\mathcal{P}}) =
    \mu(\goth g_{\mathcal{R}}\mid \goth g_{\mathcal{L}})\,.
    \label{RP2} 
  \end{equation}
  Thence, for every $f$ that is determined on
  $\mathcal{G}_{\mathcal{R}}$ we have
  \begin{equation}
    \begin{split}
      \langle f\;\vartheta_{\!\mathcal{P}}f \rangle_{\mu}& =
      \sum_{\goth g_{\mathcal{P}}}\mu(\goth g_{\mathcal{P}}) \langle
      f\mid \goth g_{\mathcal{P}} \rangle_{\mu_{\mathcal{R}}}
      \langle\vartheta_{\!\mathcal{P}}f\mid \goth g_{\mathcal{P}}
      \rangle_{\mu_{\mathcal{L}}} \\ & =\sum_{\goth
        g_{\mathcal{P}}}\mu(\goth g_{\mathcal{P}}) \langle f\mid \goth
      g_{\mathcal{P}} \rangle_{\mu_{\mathcal{R}}}^{2}
    \end{split}
    \label{RP3} 
  \end{equation}
  which cannot be negative.  Similarly we get that $\langle h
  \;\vartheta_{\!\mathcal{P}} f \rangle_{\mu} = \langle f
  \;\vartheta_{\!\mathcal{P}} h \rangle_{\mu}$.\qed
\end{pf*}

One of the important consequences of reflection positivity is a
Cauchy--Schwartz-type inequality:
\begin{equation}
  \langle f \;\vartheta_{\!\mathcal{P}} h \rangle_{\mu} \leq
  \sqrt{\langle f \;\vartheta_{\!\mathcal{P}} f \rangle_{\mu}
    \langle h \;\vartheta_{\!\mathcal{P}} h \rangle_{\mu}}\,.
  \label{RP:CS}
\end{equation}
which in turn leads to the {\em chessboard estimates\/} to be
described below.  (The reader interested in a more detailed
description of reflection positivity is referred to the review
\cite{Shlosman:RP} and the references therein).

\subsection{Uniform exponential decay for large $n$}  

In this subsection we will consider some $n$--colour models with $n
\gg 1$ and vertex factors that are uniformly bounded above and below
independently of $n$: $0< c \leq \nu_1, \dots \nu_m \leq C$.  Examples
include the $\mathrm O(n)$-type models and the corner-cubic models
discussed in \Sref{sec:models}.  However, the face-cubic
model does not fall into this category since all the multi-coloured
vertex factors vanish.  It is no coincidence that we cannot treat
these models since, as is obvious such models have a colour-symmetry
broken phase for high enough value of bond fugacity (for brevity we
omit a formal proof).

The suppression of long contours will be established by showing that
long lines of any particular colour are exponentially rare in the
length of the line. To prove this we will need the so called
chessboard estimate which in the present context reads as follows:
\begin{prop} 
  For $x \in \Lambda$ let $\omega_1, \dots \omega_k$ denote indicator
  functions for bond events that are determined by the bonds emanating
  from the site $x$.  (The $\omega_j$ need not all be distinct.) For
  any of these $\omega_j$, cover the lattice with (multiple)
  reflections of the corresponding event and let $Z_j$ denote the
  partition function constrained so that at each site, the
  appropriately reflected event is satisfied.  Then
  \begin{equation}
    \langle \prod_{j}\omega_j(x) \rangle_{\mu} \leq \left(\frac{Z_1}Z
    \right)^{\frac 1N} \dots \left(\frac{Z_k}Z \right)^{\frac 1N}.
    \label{chess-board}
  \end{equation}
\end{prop}

\begin{pf*}{Proof.}
  See Section 2.4 of Ref.~\cite{Shlosman:RP}.
\end{pf*}
Our principal result of this subsection:
\begin{thm} 
  \label{Theorem_4_3}
  Consider an $n$-colour loop model as described by \eref{part} on the
  torus $\Lambda$ (which is taken to be ``sufficiently large'') and
  suppose that the vertex factors are bounded below by $c > 0$ and
  above by $C < \infty$ uniformly in $n$.  For sites $x$ and $y$ in
  $\Lambda$, let $\mathcal L_{x,y}$ denote the probability that these
  sites belong to the same loop.  Then, provided $n$ is sufficiently
  large, there is a $\xi_{n} > 0$ such that for all values of $R$,
  \begin{equation}
    {\mathcal L}_{x,y} \leq  K\rme^{-\abs{x - y}/\xi_n}
  \end{equation}
  where $\abs{x-y}$ denotes the minimum length of a walk between $x$ and $y$
  and $K$ is a constant.
\end{thm}
\begin{rmk} For conceptual
  clarity, we will start with a proof of the case $d = 2$; all of the
  essential ideas are contained in this case.  The problems in $d > 2$
  involve some minor technicalities and the general proof can be
  omitted on a preliminary reading.
\end{rmk}   
\begin{pf*}{Proof of Theorem~\ref{Theorem_4_3} ($d = 2$).}
  Let us focus on a particular colour -- red -- and show the statement
  is true for red loops; this only amounts to a factor of $n$ in the
  prefactor. We define the ``red event'' as an event where at least
  two red bonds are connected to the site in question.  It is clear
  that there are two main types of red events: those where the red
  bonds attached to a given site line up along the straight line and
  those where they form the right angle.  These two types are shown in
  Figures~\ref{reflection}(I-a) and \ref{reflection}(II-a)
  respectively.
  \begin{figure}[hbt]    
    \begin{center}
      \includegraphics[width=14cm]{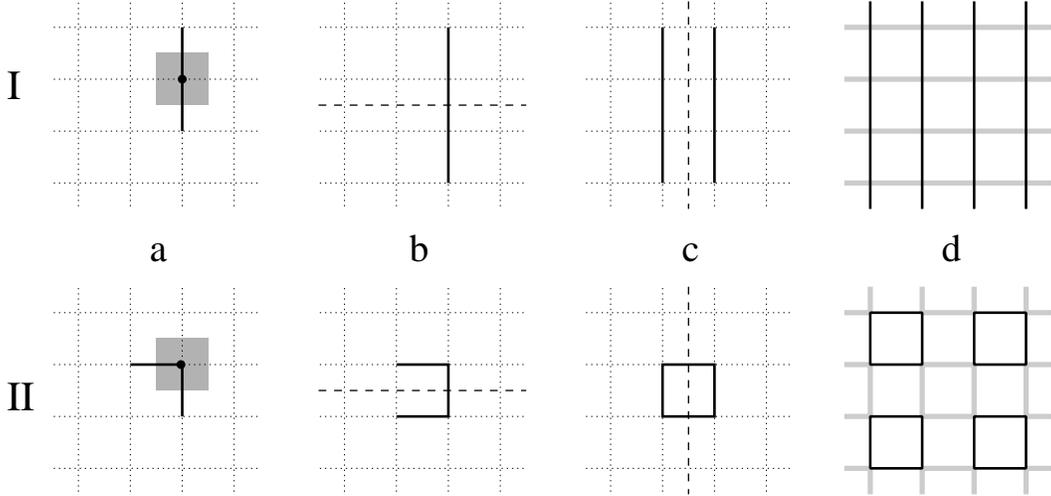}
      \caption{
        Chess-board estimate on the ``red'' events of types I and II.
        The events are shaded in (a). Parts (b) and (c) represent the
        results of the first two reflections with respect to the
        dashed lines.  The resulting tilings of the entire plane
        (torus) are shown in part (d).}
      \label{reflection}
    \end{center}
  \end{figure}
  We will denote by $\omega_{I}^{(\alpha)}$, $\alpha = 1,2$ the events
  of the first type and $\omega_{II}^{(\beta)}$, $\beta = 1, \dots 4$
  those of the second type.  Obviously there are only two distinct
  constrained partition functions which we respectively denote by
  $Z_{I}$ and $Z_{II}$.  \Fref{reflection} shows these single events
  (a), the results of the first two reflections with respect to the
  dashed lines (b), (c) and finally the configurations obtained by
  applying each reflection $\ell d =\log_2N$ times in order to
  completely tile the surface of the torus.  The grey lines correspond
  to yet unidentified bonds -- these are the degrees of freedom left
  after the process of tiling has been completed.
  
  Let us perform the estimate on $Z_{II}/Z$ first.  We claim that if
  $\goth r_{II}$ is any legitimate configuration that contributes to
  $Z_{II}$ each of the red squares -- of which there are $N/4$ -- can
  be independently replaced by vacant bonds or a square loop of any
  other colour or left as red.  Of course this may cost us an exchange
  of the ``best'' for the ``worst'' vertex factor but even so, the
  result is
  \begin{equation}
    \frac{Z_{II}}Z \leq
    \left( \frac{R^4}{1 + nR^4} \right)^{\frac N4}
    \left( \frac Cc \right)^{N}
    \leq \left ( \frac1{n^{\frac 14}}\frac Cc \right )^N.
    \label{eq:Z_II}
  \end{equation}
  
  Let us now turn attention to the $Z_I$ estimate.  We start with a
  factor of $R^N$ for the red bonds already in place -- as well as
  another worst case scenario of $C^N$.  As for the lines that are
  orthogonal to (horizontal in \fref{reflection}(I-d)) once started in
  any colour they must continue until they wrap the torus, a total
  length of $L = \sqrt N$.  There are $n$ possible choices of colour
  for each line as well as the possibility of no bonds at all.  Since
  there are a total of $L$ lines altogether, this gives
  \begin{equation}
    Z_I \leq C^N R^N(1 + nR^{L})^{L}.
    \label{eq:Z_I}
  \end{equation}
  
  To obtain our estimate on $Z$, we simply pick the even (or odd)
  sublattice of dual sites and surround each site with one of $n$
  coloured elementary loops or with a ``loop'' of vacant bonds.
  Folding in the worst case scenario for the vertex factors this gives
  \begin{equation}
    Z \geq c^N(1 + nR^4)^{\frac{N}{2}}.
    \label{partition_bound}
  \end{equation}
  The ratio may be expressed as a product of two terms namely $[R/(1 +
  nR^4)^{1/4}]^N$ and $(1 + nR^{L})^{L}/(1 + nR^4)^{N/4}$ -- times an
  additional $(C/c)^N$.  Clearly the first term is bounded by
  $n^{-N/4}$.  As for the second ratio, if $R < 1$ we may neglect
  $nR^{L}$ for $N \gg 1$ and the ratio is bounded by one.  On the
  other hand, if $R > 1$, we may neglect the 1 and we get, modulo a
  factor of $n^{L}$, another $n^{-N/4}$; we will settle for the bound
  of 1.

  Thus we have
  \begin{equation}
    \lim_{N \to \infty}\left (\frac {Z_{I}} Z \right )^{\frac 1N} \leq
    \left ( \frac1{n^{\frac 14}}\frac Cc \right )
    \label{eq:therm_limit}
  \end{equation}
  with the same upper bound for $(\frac {Z_{II}} Z )^{1/N}$ valid for
  all $N$.  We thus denote the mutual upper bound by $\epsilon_n \sim
  n^{-1/4}$.  The desired result now follows from a standard Peierls
  argument: If $x$ and $y$ are part of the same loop, some subset of
  this loop must be a self-avoiding walk of length at least
  $2\abs{x-y}$.  We enumerate all such walks and use the chess board
  estimate on each particular walk.  Then if $\epsilon_n \lambda_2 <
  1$ where $\lambda_2$ is the two dimensional connectivity constant we
  write
  \begin{equation}
    \epsilon_n \lambda_2 = \exp\{-1/2\xi_n\}
  \end{equation}
  (the factor of 2 because we must go there and back) and the stated
  result follows. \qed
\end{pf*}

\begin{pf*}{Proof of Theorem~\ref{Theorem_4_3} ($d>2$).} 
  The preliminary steps are the same as the two-dimensional case:
  There are again only two types of $\omega$'s (but with more indices)
  and two constrained partition functions which we again denote by
  $Z_I$ and $Z_{II}$.  The pattern for $Z_{II}$ is the two-dimensional
  pattern in \fref{reflection}(II-d) reflected in all directions
  orthogonal to the plane visualised.  Noting that each reflection
  doubles the number of red squares, we see that in the $Z_{II}$
  patterns there are a total of $(2^{\ell})^{d-2}\times\frac 14 L^{2}
  = \frac 14 N$ squares altogether.  Repeating the argument leading to
  \eref{eq:Z_II} we end up with exactly the same bound.  What is a
  little harder is the estimates on $Z_I$ and the partition function
  itself.

  We first claim that $Z$ can be estimated by
  \begin{equation}
    Z \geq c^N(1 + nR^4)^{\frac d4 N}.
    \label{eq:Z_bound}
  \end{equation}
  
  To achieve this, we assert that the following holds: There is a set
  of plaquettes of the lattice with the property that each bond of the
  lattice belongs to exactly one plaquette.  Once this claim is
  established it is clear that \eref{eq:Z_bound} holds; indeed there
  are just $[1/4] \times Nd$ plaquettes in question, we consider those
  configurations in which each of them is independently left vacant or
  traversed with an elementary loop in one of the $n$ possible
  colours.  Let us turn to a proof of the above assertion.
  
  Let $\mathbf{\hat e}_1, \dots \mathbf{\hat e}_d$ denote the
  elementary unit vectors.  We adapt the following notation for
  plaquettes: If, starting at $\mathbf{x} \in \mathbb{Z}^d$ we first
  move in the $\mathbf{\hat e}_j$ direction then in the $\mathbf{\hat
    e}_k$ direction and then complete the circuit we will denote this
  plaquette by $[\mathbf{\hat e}_j \diamond \mathbf{\hat
    e}_k]_\mathbf{x}$ In general we can have $[\pm \mathbf{\hat e}_j
  \diamond \pm \mathbf{\hat e}_k]_\mathbf{x}$ and it is noted that
  $[ \mathbf{\hat e}_j \diamond \mathbf{\hat e}_k]_\mathbf{x}$
  = $[ \mathbf{\hat e}_k \diamond \mathbf{\hat
    e}_j]_\mathbf{x}$
  
  Starting with the origin, consider the following list of
  instructions for plaquettes:
  \begin{equation}
    [\mathbf{\hat e}_1 \diamond -\mathbf{\hat e}_2]_{0}, \
    [\mathbf{\hat e}_2 \diamond -\mathbf{\hat e}_3]_{0}, \dots ,
    [\mathbf{\hat e}_d \diamond -\mathbf{\hat e}_1]_{0}.
    \label{eq:plaquette_instructions}
  \end{equation}
  So far so good -- each bond emanating from the origin belongs to
  exactly one plaquette.  If $\mathbf{x} = (x_1, \dots x_d)$ let us
  define $\pi_j(\mathbf{x}) = (-1)^{x_j}$ to be the parity of the
  $j^{\mbox{th}}$ coordinate.  Then at the site $\mathbf{x}$, select
  the following: $[\pi_1(\mathbf{x})\mathbf{\hat e}_1 \diamond
  -\pi_2(\mathbf{x})\mathbf{\hat e}_2]_\mathbf{x}, \dots
  [\pi_d(\mathbf{x})\mathbf{\hat e}_d \diamond
  -\pi_1(\mathbf{x})\mathbf{\hat e}_1]_\mathbf{x}$.  Taking the union
  of all these lists, it is clear that indeed each bond belongs to
  {\em at least\/} one plaquette, the only question is whether there
  has been an over-counting.
  
  To settle this issue we establish the following assertion: Let
  $\mathbf{x}$ denote a site.  Then a plaquette is specified by the
  instructions at $\mathbf{x}$ if and only if the same plaquette is
  specified by the neighbours of $\mathbf{x}$ on that plaquette.
  
  Indeed, consider one such plaquette namely $[\pi_j(\mathbf{x})
  \mathbf{\hat e}_j \diamond -\pi_{j+1}(\mathbf{x})\mathbf{\hat
    e}_{j+1}]_\mathbf{x}$.  (If necessary we use the convention
  $\mathbf{\hat e}_{d+1} = \mathbf{\hat e}_1$). One of the relevant
  neighbours is $\mathbf{x^\prime} = \mathbf{x} +
  \pi_j(\mathbf{x})\mathbf{\hat e}_j$.  But at $\mathbf{x^\prime}$, we
  have instructions for the plaquette
  $[\pi_j(\mathbf{x^\prime})\mathbf{\hat e}_j \diamond
  -\pi_{j+1}(\mathbf{x^\prime})\mathbf{\hat
    e}_{j+1}]_\mathbf{x^\prime}$.  Since $\pi_j(\mathbf{x^\prime}) =
  -\pi_j(\mathbf{x^\prime})$ and $\pi_{j+1}(\mathbf{x^\prime}) =
  \pi_{j+1}(\mathbf{x^\prime})$ this is seen to be the same plaquette.
  The other neighbour, at $\mathbf{\tilde x} = \mathbf{x} -
  \pi_{j+1}\mathbf{\hat e}_{j+1}$, follows from the same argument.
  Reversing the r\^oles of $\mathbf{x}$ and $\mathbf{x^\prime}$ (as
  well as $\mathbf{x}$ and $\mathbf{\tilde x}$) establishes the `only
  if' part of the assertion.
  
  It is thus evident that no bond can belong to two plaquettes: If
  $\mathbf{x}$ and $\mathbf{x^\prime}$ are neighbours and there are
  instructions coming from $\mathbf{x}$ to make some particular
  plaquette that includes the bond $\langle
  \mathbf{x},\mathbf{x^\prime} \rangle$ then:

  \noindent (a)  These are the only instructions coming from $\mathbf{x}$ 
  pertaining to this bond.

  \noindent (b)  The list at $\mathbf{x^\prime}$ (and the other corners of 
  this plaquette) also include this plaquette.

  \noindent (c)  If $\mathbf{y}$ is another neighbour of $\mathbf{x}$ 
  which is {\em not\/} in this plaquette there cannot be instructions
  at $\mathbf{y}$ to include the plaquette containing $\mathbf{x}$,
  $\mathbf{x^\prime}$ and $\mathbf{y}$.
  
  Items (a) and (b) are immediate.  To see item (c), note that by the
  assertion, such instructions would also have to appear on the list
  at $x$.  But, by item (a), they don't.  \Eref{eq:Z_bound} is now
  established.
  
  Finally, let us estimate $Z_{I}$ from above.  First it is noted that
  the $Z_{I}$ patterns consist of parallel lines (pointing in the
  direction of the original red bonds) that pass through every site.
  Hence the only loops we can draw are confined to the various
  $(d-1)$--dimensional hyperplanes orthogonal to these lines.  Modulo
  vertex factors -- which we estimate by $C^{N}$ -- each hyperplane
  yields the $(d-1)$--dimensional partition function.  We thus have
  \begin{equation}
    Z_{II}  \leq  R^N C^N (\Xi)^{L}
    \label{eq:Z_II_d}
  \end{equation}
  where $\Xi$ is the $(d-1)$--dimensional partition function with all
  vertex factors equal to unity.  (This observation is not of crucial
  importance -- the key ingredient is the number of available bonds --
  but it serves to compartmentalise the argument.)  Let us tend to the
  estimate of $\Xi$.
  
  We denote by $\Xi_M$ the contribution to $\Xi$ that comes about when
  there are exactly $M$ bonds in the configuration. Given the
  placement of the bonds, let us count (estimate) the number of ways
  that they can be organised into self-returning walks and then
  independently colour each walk.  Throwing in a combinatoric factor
  for the placement of the $M$ bonds and the fact that there cannot
  conceivably be more than $M/4$ loops to colour we arrive at
  \begin{equation}
    \Xi_M  \leq {{(d-1)L^{d-1}}\choose M} R^{M}n^{\frac M4}\times Y(M)
    \label{eq:Xi_bound}
  \end{equation}
  where $Y(M)$ is defined as follows: For $M$ bonds, consider their
  placement on the lattice chosen so as to maximise the number of ways
  in which the resulting (colourless) graph can be decomposed into
  distinct loops. (Such decompositions are done by ``resolving'' all
  intersecting vertices into connected pairs of bonds, similar to
  \fref{O(n)_vertices}.) Then the quantity $Y(M)$ denotes this maximum
  possible number of decompositions.
  
  Let $\goth z$ (which equals $2(d-1)-1=2d -3$) denote the greatest
  possible number of local options for an on-going self-avoiding walk.
  We claim that
  \begin{equation}
    Y(M) \leq (2\goth z)^M.
    \label{eq:Y_bound}
  \end{equation}
  Indeed, take this optimal placement of $M$ bonds and, starting at
  some predetermined bond (and moving in some predetermined direction)
  draw a loop of length $P$.  There are no more than $\goth z^P$
  possibilities for this loop.  Then what is left over cannot yield a
  total better than $Y(M - P)$. This must be done for every possible
  value of $P$ and summed.  Defining $Y(0) = 1$, we arrive at the
  recursive inequality
  \begin{equation}
    Y(M) \leq \sum_{0 < P \leq M}\goth z^PY(M-P)
  \end{equation}
  and the bound in \eref{eq:Y_bound} can be established inductively.
  Putting together equations (\ref{eq:Xi_bound}) and
  (\ref{eq:Y_bound}) we find
  \begin{equation}
    \Xi \leq \sum_M {{(d-1)L^{d-1}}\choose M} 
    R^{M}n^{\frac M4}(2\goth z)^M =
    (1 + 2\goth zRn^{\frac 14})^{(d-1)L^{d-1}}.
    \label{eq:Xi}
  \end{equation}
  And thus
  \begin{equation}
    Z_{I} \leq C^NR^{N}[(1 + 2\goth zRn^{\frac 14})^{(d-1)L^{d-1}}]^L =
    [CR(1 + 2\goth zRn^{\frac 14})^{(d-1)}]^N.
    \label{eq:Z_I_bound}
  \end{equation}
  
  An estimate for the straight segments is nearly complete.  We write
  \begin{equation}
    \left[\frac {Z_I}Z\right]^{\frac 1N} \leq
    \frac cC\frac {R(1 + 2\goth zRn^{\frac 14})^{(d-1)}}
    {(1 + nR^4)^{d/4}}\,.
    \label{eq:Z_I/Z}
  \end{equation}
  Let us split the power of the denominator: $d/4 = 1/4 + (d-1)/4$;
  the first term will handle the $R$ in the numerator and this ratio
  is less than $n^{-1/4}$.  As for what is left over, it is easy to
  see that
  \begin{equation}
    \frac{1 + 2\goth zRn^{\frac 1 4}}
    {(1 + nR^4)^{\frac 1 4}} = 
    \frac{1}{(1 + nR^4)^{\frac 1 4}} +
    2\goth z \frac{Rn^{\frac 1 4}}
    {(1 + nR^4)^{\frac 1 4}}
    \leq 1 + 2\goth z\,.
  \end{equation}
  Thus we have
  \begin{equation}
    \left[\frac {Z_I}Z\right]^{\frac 1N} \leq
    \frac Cc (1 + 2\goth z)^{d-1} \frac 1{n^{1/4}}
    \label{eq:Z_I/Z_bound}
  \end{equation}
  so again all of the $\omega$'s have estimates with the scaling of
  $n^{-1/4}$.  The remainder of the argument follows the same course as
  the two-dimensional argument with modifications where appropriate. \qed
\end{pf*}

\subsection{Translational symmetry breaking and related phase transitions}
\label{sec:transl_sym}

So far we have shown that for a sufficiently large $n$, there are no
phase transitions associated with the divergent loop correlation
length. However, this does not rule out the possibility of phase
transitions of a totally different nature -- in fact we shall see that
such transitions do indeed take place.  The type of these transitions
happens to be sensitive to the lattice structure and the vertex
factors that correspond to loop intersections. This leads to a variety
of critical phenomena which has not been observed in the case of a
honeycomb lattice \cite{Domany-81}.

We shall consider the case of all vertex factors being positive and
uniformly bounded. Let the number of colours $n$ be very large and
consider the high fugacity limit of $R \to \infty$. The model becomes
fully-packed in this limit, i.e.  every bond is occupied. But since
$n$ is large, according to the estimates of the previous subsection
the partition function is dominated by configurations in which there
are as many loops as possible.  In two dimensions, it is clear that
these ``maximum entropy states'' break the translational symmetry: the
loops of different colours may run around either odd or even
plaquettes of the lattice.  (We remark that in $d>2$, the situation is
considerably more complicated and is currently under study.)

On the other hand, if the bond fugacity is low, the system must be
translationally invariant. Therefore one would anticipate that at some
large but finite value of $R$ a translational symmetry-breaking
transition takes place.  Due to a doubly positionally degenerate
nature of the resulting high-density state, we find it natural to
expect that such transition is of the Ising type.  Indeed, it appears
to be very similar to the transition in the Ising antiferromagnet.

Let us now turn to the rigorous arguments supporting this qualitative
picture.

\begin{thm} 
  \label{thm:symmetry-breaking}
  Consider an $n$-colour loop model on $\mathbb{Z}^2$ with vertex
  factors bounded below by $c>0$ and above by $C < \infty$ uniformly
  in $n$. Then, if $n$ is sufficiently large there is a phase
  transition characterised by the breaking of translational symmetry.
  In particular, there exist $\epsilon$ (sufficiently small) and
  $\Delta$ (sufficiently large) such that for $R^4 n < \epsilon$ all
  states that emerge as limits of torus states are translation
  invariant. On the other hand, if $R^4 n > \Delta$, there are (at
  least) two states with broken translational symmetry.
\end{thm}
\begin{pf*}{Proof.}
  Let us start with the formal proof of translation-invariance at low
  fugacity.
  
  Any given site belongs to zero, two or four bonds. Let $\odot$,
  $\ominus$ and $\oplus$ denote the corresponding events and
  $Z_\odot$, $Z_\ominus$ and $Z_\oplus$ the constrained partition
  functions in which every site is of the stated type. We will
  estimate $\mathrm{Prob}(\ominus) \leq \left(Z_\ominus/Z_\odot
  \right)^{1/n}$ and $\mathrm{Prob}(\oplus) \leq \left(Z_\oplus/Z_\odot
  \right)^{1/n}$; obviously $Z_\odot = 1$. In calculating $Z_\oplus$
  we notice that in all configurations, each bond must be coloured and
  there are at most $N/2$ separate loops. This gives
  \begin{equation}
    \label{Z-four}
    Z_\oplus \leq \left[R^2 n^{1/2}\right]^N Y(2N)
  \end{equation}
  where $Y(M)$ is the same quantity that appears in \eref{eq:Xi_bound}.
  Similarly, in $Z_\ominus$ half the bonds are used (resulting in a factor
  of $R^N$) and there are no more than $N/4$ loops. We get
  \begin{equation}
    \label{Z-two}
    Z_\ominus \leq \left[R n^{1/4}\right]^N Y(N)
  \end{equation}
  -- essentially the square root of the above.
  
  In order to show that there is translation invariance it is
  sufficient to establish the following: Let $A$ and $B$ denote local
  events, let $T_{\mathbf x}$ be the translation operator to the point
  $x \in \Lambda$ and let $\mathbf{\hat e}$ denote any unit vector.
  Then we must show that for all ${\mathbf x}$ with $\abs{{\mathbf
      x}}$ large the probabilities of $\mu_\Lambda \left(A\cap
    T_{\mathbf x} B\right)$ and $\mu_\Lambda \left(A\cap T_{\mathbf x
      + \mathbf{\hat e}} B\right)$ are essentially the same. We note
  that either the supports of $A$ and $T_{\mathbf x} B$ are attached
  by a $\ast$-connected path of sites of type $Z_\ominus$ and
  $Z_\oplus$, or they are separated by a connected circuit of cites of
  type $Z_\odot$. (Two sites are considered $\ast$-connected if they
  share the same plaquette.)  However, in case the support of A is
  surrounded by a connected circuit of empty sites, it is clear that
  for any $\mathbf{\hat e}$, $A$ and $T_{\mathbf{\hat e}} A$ have the
  same probability.  Thus we arrive at
  \begin{equation}
    \label{eq:mu_difference}
    \abs{ \mu_\Lambda \left(A\cap T_{\mathbf
          x} B\right) - \mu_\Lambda \left(A\cap T_{\mathbf x +
          \mathbf{\hat e}} B\right)} \leq 
    \mu_\Lambda \left(\mathbb C_{A, T_\mathbf
        x B} \right)
  \end{equation}
  where $\mathbb C_{A, T_\mathbf x B}$ is the event that there is a
  $\ast$-connected path of non-$\odot$ sites between the support of A
  and that of $T_{\mathbf x} B$. Now, provided $R n^{1/4}$ is small,
  $\mu_\Lambda \left(\mathbb C_{A, T_\mathbf x B} \right) \leq
  \rme^{-\kappa \abs{x}}$ with $\rme^{-\kappa} \sim \alpha R n^{1/4}$
  for some constant $\alpha$.  Thence translation invariance (among
  the states that emerge from the torus) is established.
  
  Let us now turn attention to the opposite limit, namely $R n^{1/4}
  \gg 1$. We claim that with high probability the plaquettes on the
  lattice are of one of the two types: they are either surrounded by
  four bonds of the same colour, or by bonds of four different
  colours.
 
  \begin{figure}[htb]
    \begin{center}
      \includegraphics[width=5cm]{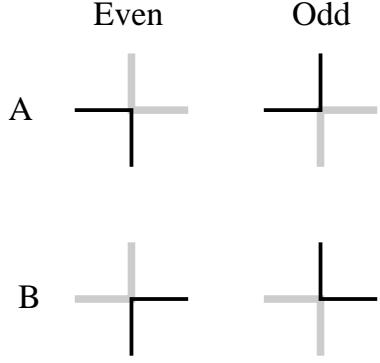}
      \caption{
        Even and odd sites of A- and B-types.}
      \label{A-B_sites}
    \end{center}
  \end{figure}
  
  We will again define site events representative of the two purported
  ground states. If a site is on the even sublattice, we say it is of
  the A-type if it has four bonds of two different colours, one colour
  occupying the positive $\mathbf{\hat e}_1$ and $\mathbf{\hat e}_2$
  directions and the other, the negative $\mathbf{\hat e}_1$ and
  $\mathbf{\hat e}_2$ directions. The odd site of the A-type is
  defined as the mirror reflection of an even A-type site. A site is
  said to be of B-type if the r\^oles of the even and the odd
  sublattices are reversed -- see \fref{A-B_sites}. Any site which is
  not of one of these two types will be called ``bad''. There are
  three choices for such bad sites, namely housing no bonds, only two
  bonds and a ``+'' configuration in which the $\mathbf{\hat e}_1$ and
  $-\mathbf{\hat e}_1$ directions have the same colour and similarly
  for the $\pm \mathbf{\hat e}_2$ directions. The first two are the
  old $\odot$ and $\ominus$ from the preceding portion of this proof,
  while the constrained partition function for the latter event,
  denoted (for the lack of better notations) by $Z_{\#}$ can be
  bounded by
  \begin{equation}
    \label{eq:Z_plus}
    Z_{\#} \leq R^{2N} n^{2 \sqrt{N}}\,.
  \end{equation}
  On the other hand, we can always estimate $Z$ from below by $\left[R
    n^{1/4}\right]^2N$ as discussed previously. Thus we have (as
  $\abs{\Lambda} \to \infty$)
  \begin{subequations}
    \label{eq:mu_estimates}
    \begin{gather}
      \mu_\Lambda(\odot) \leq R^{- 2} n^{- 1/2}\,, \\
      \mu_\Lambda(\ominus) \leq n^{- 1/2}\,, \\
      \mu_\Lambda({\#}) \leq n^{- 1/4}\,.
    \end{gather}
  \end{subequations}
  Thus almost all sites are either A-type or B-type. However, there is
  no constraint that A and B sites cannot be neighbours. But, as we
  show below, this possibility is also suppressed for $n \gg 1$.
  Firstly, let us notice that the reflections through bonds map the
  even into the odd sites (and vice versa). By definition, the A-even
  and A-odd events are mirror images, and similarly for B. Thus, under
  multiple reflections, A-sites map to A-sites and B's to B's. Hence,
  if (A-B) is the event that the origin is of the A-type and its right
  nearest neighbour is of the B-type, we have
  \begin{equation}
    \label{eq:mu_AB}
    \mu_\Lambda(\mbox{A-B}) \leq 
    \left(\frac{Z_{\mbox{\scriptsize A-B}}}{Z}\right)^{2/N}
  \end{equation}
  where $Z_{\mbox{\scriptsize A-B}}$ is the partition function
  constrained according to the pattern shown in \fref{A-B_pattern}.
  \begin{figure}[hbt]
    \begin{center}
      \includegraphics[width=7cm]{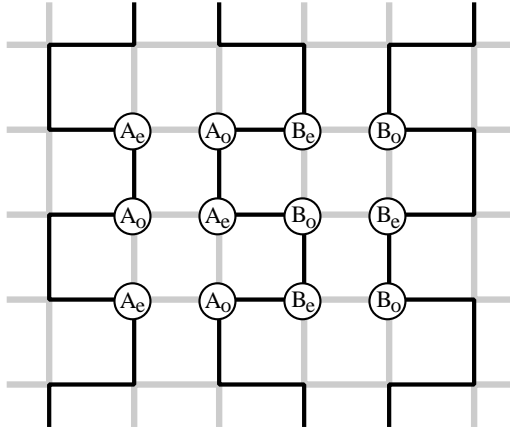}
      \caption{
        Tiling of the torus obtained by multiple reflections of the
        (A-B) event.  The sites marked as $\mathrm{A_e}$ are the even
        sites of the A-type etc. The zigzag lines that form loops
        wrapping the torus and separating the columns of A- and
        B-sites are shown black.}
      \label{A-B_pattern}
    \end{center}
  \end{figure}
  Now, the interior of the A or B columns still consists of the
  four-bond loops -- just as a pure A or pure B-type tiling would.
  However, as is easily seen in \fref{A-B_pattern}, a boundary between
  the columns is formed by a zigzag line of a single colour running
  around the torus. Thus we arrive at
  \begin{equation}
    \label{eq:Z_AB}
    Z_{\mbox{\scriptsize A-B}} = R^{2N} n^{\frac{1}{4} N} 
    n^{\frac{1}{2} \sqrt{N}}
  \end{equation}
  and hence, in the large $N$ limit, $\mu_\Lambda(\mbox{A-B}) \leq
  n^{-1/2}$.
  
  The rest of the proof follows easily: A connected cluster of A-type
  sites has, on its boundary a ``bad'' site or an A-B pair; similarly
  for the clusters of B-sites. Thus, if $R n ^{1/4}$ and $n$ are large
  enough, the possibility that any given site is isolated from the
  origin is small. Thence, given that the origin is of the A-type --
  which has probability very near one half -- the population of B-type
  sites is small and vice versa. Evidently, there are two states, one
  with an abundance of A-sites and the other with an abundance of B's.
  It is not hard to see that in these two states, translation symmetry
  is broken -- both are dominated by the appropriate staggered
  patterns. \qed
\end{pf*}

\section{Final remarks and conclusions}

So far we have succeeded in proving the following general statements:
\begin{itemize}
\item {A generic multi-coloured loop model with {\em all\/} vertex
    factors $\nu_1, \ldots \nu_m$ uniformly bounded from above and
    below does not have a phase transition corresponding to the
    divergence of the loop size in {\em any\/} dimension {\em
      provided\/} that the number of colours $n$ is sufficiently
    large.}
\item {In two dimensions these models undergo a different phase
    transition, presumably of the Ising type, that is associated with
    breaking the {\em translational\/} symmetry\footnote{ At present
      it is not clear whether this statement holds in $d>2$. While the
      ``ground states'' (the states at $R=\infty$) of the system are
      {\em not\/} translationally invariant, their degeneracy appears
      to grow very fast with $d$, and it is not obvious that the
      resulting ``entropy'' will not destroy the transition at any
      finite value of $R$.}. While the examples of the models that
    have been considered in this paper all have their vertex factors
    independent of the particular arrangements of different colours
    entering the vertex, all these results remain valid even if this
    is not the case as long as all different vertex factors are still
    reflection-symmetric and bounded from above and below by some
    positive numbers.}
\end{itemize}

{A clear example of a model that does not follow this rule is the loop
  model derived from the face-cubic spin model in \sref{sec:face-cubic}.
  Here the bonds of different colours are simply not allowed to share a
  vertex, making the corresponding vertex factor vanish.  It is clear
  that as the bond fugacity $R$ increases, the system will have a
  phase transition (possibly of the first order) associated with
  breaking the {\em colour\/} symmetry. Indeed, the only way to ``pack''
  more bonds into the system is to force all of them to be of the same
  colour. This transition is very similar to the Widom-Rowlinson
  transition.}

{Another interesting observation can be made about the two-dimensional
  loop model with all two- and four-leg vertices having the same
  weight (this is the model originating from the corner-cubic spin
  model -- see \sref{sec:corner-cubic}). Comparing the states of the
  system at $R=\infty$ when all bonds are occupied (the fully packed
  limit), and at $R=1$ when no additional weight is associated with
  placing extra loops, we conclude that the $(n+1)$-coloured model at
  $R=\infty$ is identical to the $n$-coloured model at $R=1$. Indeed,
  one should simply consider the vacant bonds at $R=1$ as being
  coloured grey to turn the $n$-coloured loop system into a
  fully-packed $(n+1)$-coloured system.
    
  Now start with the large enough $n$ at $R=\infty$. The result of
  \sref{sec:transl_sym} guarantees the existence of a broken
  translational symmetry in this case. By the argument presented here
  this means the existence of a broken symmetry in $(n-1)$-coloured,
  $R=1$ case. Now let us continuously increase the value of the bond
  fugacity in this $(n-1)$-coloured model. As $R$ grows, there are
  only two possible scenarios: the broken translational symmetry is
  either lost via a phase transition or is retained all the way to
  $R=\infty$. The former case corresponds to the {\em intermediate\/}
  symmetry-broken phase surrounded by the phase transitions at $R \leq
  1$ and $R \geq 1$. If the latter scenario is realised, then go to
  $R=\infty$ and repeat the process of mapping it onto a lower $n$
  model. Notice that this latter scenario can not continue all the way
  to $n=1$ because the $n=1$, $R=1$ case is nothing but a loop
  representation for the Ising magnet at $T=0$ which does not have
  broken translational symmetry.\footnote{Indeed, these loops
    appearing in the $T=0$ ``high temperature'' expansion are also the
    domain walls of the Ising model on the dual lattice at $T \to
    \infty$. These dual spins are assigned their values independently
    with probability $1/2$, and the contours separating regions of
    opposite type are manifestly translation invariant.} Therefore one
  {\em must\/} find an intermediate phase at some not very large value
  of $n$ (although we cannot rule out a possibility of it being just the
  point $R=1$).}
  
{From the above we also learn that the $n=2$ (Ashkin--Teller-like)
  model is {\em not\/} critical in its fully-packed limit (since it is
  just the $T=0$ Ising model). On the other hand the $\mathrm O(2)$
  loop model discussed in \sref{sec:On} {\em is\/} critical in this
  limit (it maps onto the square ice model). But the only difference
  between the two is the factor of 3 in the same-colour four-leg
  vertex factor.}
  
On the basis of this observation (along with the first two remarks of
this section) we reiterate that the vertex factors and lattice details
are often important for determining the phase diagram of a particular
model.

\appendix 
\section{Vertex factor for the $\mathrm O(n)$ vertices and loop decomposition}
\label{sec:KMF}

In this appendix we present the calculation of the generic vertex
factor appearing in the loop expansion of the $\mathrm O(n)$-type model
featured in \ref{sec:On}. As it turns out, this calculation is
independent of the lattice details (such as the dimensionality or
coordination number) which is reflected in our notation. Thus let
$\Omega_n$ denote the $n$-dimensional solid angle and let $\mathbf{ S}
= \left(S_1, \ldots S_n \right)$ denote the components of a
$n$-dimensional unit vector. Since we only consider the spin at one
particular site, the actual site index is conveniently dropped for
now. The vertex factor we need to calculate is
\begin{equation}
  \nu_{m_1, \ldots m_n} = \frac{1}{\abs{\Omega_n}}
  \int_{\Omega_n} S_1^{2m_1} \ldots S_n^{2m_n} \,\rmd \Omega_n 
  \label{eq:vertex_factor} 
\end{equation}
where $\abs{\Omega_n} = {2 \pi^{n/2}}{[\Gamma \left(n/2 \right)]^{-1}} $
is the total solid angle and $\rmd \Omega_n /\abs{\Omega_n}$ is the
Haar measure.  The object $\nu_{m_1, \ldots m_n}$ is, of
course, the ``vertex factor'' originating from $2m_1$ terms from the
first component, \ldots $2m_n$ terms from the $n$-th component
associated with these numbers of these types of bonds entering into a
vertex.

\begin{prop} 
  \label{Proposition_A1}
  Let $\phi(2m)$ -- which is equal to $(2m-1)!!$ -- denote the number
  of different ways of dividing $2m$ objects into $m$ distinct pairs.
  Then
  \begin{equation}
    \nu_{m_1, \ldots m_n} = D_n(M) \prod_{i: \; m_i \neq 0} \phi(2m_i)
    \label{eq:proposition_A1}
  \end{equation}
  where $M = \sum_{i=1}^{n} m_i$ and 
  \begin{equation}
    D_n(M) = 2^{-M} \Gamma{\left(\frac{n}{2}\right)}/ 
    \Gamma{\left(M + \frac{n}{2}\right)}
    = (n-2)!!/(n+2M-2)!!\,. 
  \end{equation}
\end{prop}
\begin{rmk}
  The calculation of the value for $\nu_{m_1, \ldots m_n}$ has
  appeared before, {\em cf.}\ Ref.~\cite{MS:self-avoiding} and
  references therein.  However, the interpretation that emerges from
  this formula has, to our knowledge, not been discussed previously
  ({\em cf.}\ the remark following the proof).
\end{rmk}
\begin{pf*}{Proof.}
  The proof boils down to the calculation of the above integral.
  We write
  \begin{equation}
    \rmd \Omega_n = 
    \sin^{n-2}{\theta_{n}} \, \rmd {\theta_{n}} \, \rmd \Omega_{n-1}
    \label{eq:integral_recursion}  
  \end{equation}
  where ${\theta_{n}} = \mathbf{ S}\cdot \hat\mathbf{ e}_{n}$. Also,
  \begin{equation}
    \mathbf{S} = \left(\mathbf{T}\sin{\theta_{n}},\, \cos{\theta_{n}}\right)
    \label{eq:spin_recursion}  
  \end{equation}
  where $\mathbf{T}$ is an $(n-1)\,$-dimensional unit vector. Finally, we
  note the fact that
  \begin{equation}
    \abs{\Omega_n }  = \abs{\Omega_{n-1}} 
    \frac{\sqrt{\pi}\,\Gamma{\left(\frac{n}{2}-1\right)}}
    {\Gamma{\left(\frac{n}{2}\right)}}
    \label{eq:solid_angle}  
  \end{equation}
  
  Starting with the definition \eref{eq:vertex_factor} with the help of
  equations (\ref{eq:integral_recursion})--(\ref{eq:solid_angle}), the
  recursion relation is obtained:
  \begin{equation}
    \begin{split}
      \nu_{m_1, \ldots m_n}& = \frac{1}{\abs{\Omega_n}}
      \int_{\Omega_n} S_1^{2m_1} \ldots S_n^{2m_n} \, \rmd \Omega_n
      \\
      { }& = \frac{1}{\abs{\Omega_{n-1}}} \int_{\Omega_{n-1}}
      T_1^{2m_1} \ldots T_{n-1}^{2m_n} \, \rmd \Omega_{n-1}
      \\
      & {\quad}\times \, \frac{\Gamma{\left(\frac{n}{2}\right)}}
      {\sqrt{\pi}\,\Gamma{\left(\frac{n}{2}-1\right)}} \int
      \left(\sin{\theta_{n}}\right)^{2 \sum_{i=1}^{n-1}m_i+n-2}
      \left(\cos{\theta_{n}}\right)^{2m_n}\, \rmd {\theta_{n}}
      \\
      { }& = \nu_{m_1, \ldots m_{n-1}} \frac{\Gamma{\left(m_n
            +\frac{1}{2}\right)}\,G_{n-1}}{\sqrt{\pi}\;G_{n}}\,,
  \end{split}
  \label{eq:vertex_recursion}  
\end{equation}
  where
  \begin{equation}
    G_{k} = \frac{\Gamma{\left(2\sum_{i=1}^k {m_i} +\frac{k}{2}\right)}}
    {\Gamma{\left(\frac{k}{2}\right)}}\,.
    \label{eq:G_def}  
  \end{equation}
  
  Using the recursion relation \ref{eq:vertex_recursion} to reduce the
  dimensionality of the spin space to 1 (and the fact that $\nu_{m_1} =
  1$), we arrive to the desired expression:
  \begin{equation}
    \begin{split}
      \nu_{m_1, \ldots m_n} & = \frac{\Gamma{\left(
            \frac{n}{2}\right)}} {\pi^{n/2}\,\Gamma{\left(M +
            \frac{n}{2}\right)}} \; \prod_{i=1}^n \Gamma{\left( {m_i}
          +\frac{1}{2}\right)}
      \\
      { }& = \frac{\Gamma{\left( \frac{n}{2}\right)}}
      {2^{M}\,\Gamma{\left(M + \frac{n}{2}\right)}} \; \prod_{i: \;
        m_i \neq 0} (2 m_i - 1)!!\;.
    \end{split}
    \label{eq:vertex_a}  
  \end{equation}
\qed
\end{pf*}
\begin{rmk}
  As a consequence of the expression~(\ref{eq:proposition_A1}) we may
  decompose the coloured bond configurations to obtain a colourless
  loop model (which generalises the result of \eref{O(n)_loops}
  obtained for the square lattice). In particular, at each vertex we
  have a factor which depends only on the total number of bonds
  entering the vertex (and the dimensionality of the spin $n$) times
  the number of ways that the bonds of the various colours can be
  paired. By choosing a particular pairing scheme at each vertex, the
  bond configuration breaks, unambiguously, into a collection of
  self-returning walks (loops). Configurations with the same set of
  loops -- that is to say the same set of bonds and the same pairing
  schemes (walking instructions) at each vertex -- but which differ in
  the colours of the loops are seen to have identical weights. Thus we
  consider configurations $\mathcal{K}$ of loops (bonds + walking
  instructions) and the partition function becomes the analog of the
  expression \ref{O(n)_loops}:
  \begin{equation}
    Z = \sum_{\mathcal{K}} \left(\frac{\lambda}{n}\right)^{b(\mathcal{K})}
    \nu_1^{m_1} \ldots  \nu_k^{m_k}
    n^{\ell(\mathcal{K})}
    \label{O(n)_loops_gen}
  \end{equation}  
  with $\nu_p$ being the factor for a vertex with $2p$ bonds given by
  \begin{equation}
    \label{eq:vert_factor}
    \nu_p = \nu_p(n) = \frac{n^{p}\,\Gamma{\left( \frac{n}{2}\right)}}
    {2^{p}\,\Gamma{\left(p + \frac{n}{2}\right)}} 
    =  \frac{n^{p}}
    {n (n+2) \ldots (n+2p-2)} \,.
  \end{equation}
  
  We observe that the right hand side of \eref{O(n)_loops_gen}
  provides a well defined model for all $n$ and can be continued --
  essentially with no ambiguity -- to non-integer values.  As usual,
  we may {\em define\/} correlation functions via loop probabilities
  or be content with the numerators containing strings divided by
  denominators without.
  
  In the previous discussions of these issues (see
  Ref.~\cite{MS:self-avoiding} and references therein) the goal has
  been to derive the self-avoiding walk as the $n \to 0$ limit of the
  $\mathrm O(n)$ models. Thus the vertex weights themselves get
  continued without reference to intermediate models. (And indeed, the
  self-avoiding walk does emerge as $n\to 0$.)  Although we have only
  treated the case where the original spin models were defined by $Z =
  \mathrm{Tr}\prod_{\langle i, j \rangle} \left(1 + \lambda
    \mathbf{S}_i \cdot \mathbf{S}_j \right)$ as opposed to $Z =
  \mathrm{Tr}\prod_{\langle i, j \rangle} \exp \left(\lambda
    \mathbf{S}_i \cdot \mathbf{S}_j \right)$ -- we claim that the
  latter case also leads to graphical models that are well defined for
  non-integer $n$. (The derivation is somewhat intricate and will
  eventually appear in a future publication.) However, we remark
  that in either case, the $n \to 0$ limit is, in this context, quite
  simple to understand.
\end{rmk}

\ack The authors would like to thank J.~Kondev, R.~Kotecky and
T.~Spencer for the interesting discussions. LC and KS were supported
in part by NSA Grant No.\ MDA904-98-1-0518 and NFS Grant No.\ 
99-71016; LPP was supported in part by DOE Grant No.\ 
DE-FG02-90ER40542.


\end{document}